\documentclass[aps,prl,preprint,showpacs,superscriptaddress,floatfix]{revtex4-1}

\usepackage{graphicx}

\usepackage{bm}        
\usepackage{epstopdf}

\usepackage{subfig}
\usepackage{amsmath}	
\usepackage{amssymb}    

\newcommand{\be}{\begin{eqnarray}}
 \newcommand{\ee}{\end{eqnarray}}

\begin{document}



\title{Resonantly enhanced pair production in a simple diatomic model}

\author{Fran\c{c}ois Fillion-Gourdeau}
\email{filliong@CRM.UMontreal.ca}
\affiliation{Centre de Recherches Math\'{e}matiques, Universit\'{e} de Montr\'{e}al, Montr\'{e}al, Canada, H3T 1J4}
\altaffiliation[Also at ]{School of Mathematics and Statistics, Carleton University, Ottawa, Canada, K1S 5B6; and \\Fields Institute, University of Toronto, Toronto, Canada, M5T 3J1}

\author{Emmanuel Lorin}
\email{elorin@math.carleton.ca}
\affiliation{School of Mathematics and Statistics, Carleton University, Ottawa, Canada, K1S 5B6}
\altaffiliation[Also at ]{Centre de Recherches Math\'{e}matiques, Universit\'{e} de Montr\'{e}al, Montr\'{e}al, Canada, H3T 1J4}

\author{Andr\'{e} D. Bandrauk}
\email{Andre.Dieter.Bandrauk@USherbrooke.ca}
\affiliation{Laboratoire de chimie th\'{e}orique, Facult\'{e} des Sciences, Universit\'{e} de Sherbrooke, Sherbrooke, Canada, J1K 2R1}
\altaffiliation[Also at ]{Centre de Recherches Math\'{e}matiques, Universit\'{e} de Montr\'{e}al, Montr\'{e}al, Canada, H3T 1J4}
\date{\today}

\begin{abstract}
A new mechanism for the production of electron-positron pairs from the interaction of a laser field and a fully stripped diatomic molecule in the tunneling regime is presented. When the laser field is turned off, the Dirac operator has resonances in both the positive and the negative energy continua while bound states are in the mass gap. When this system is immersed in a strong laser field, the resonances move in the complex energy plane: the negative energy resonances are pushed to higher energies while the bound states are Stark shifted. It is argued here that there is a pair production enhancement at the crossing of resonances by looking at a simple 1-D model: the nuclei are modeled simply by Dirac delta potential wells while the laser field is assumed to be static and of finite spatial extent. The average rate for the number of electron-positron pairs produced is evaluated and the results are compared to the single nucleus and to the free cases. It is shown that positrons are produced by the Resonantly Enhanced Pair Production (REPP) mechanism, which is analogous to the resonantly enhanced ionization of molecular physics. This phenomenon could be used to increase the number of pairs produced at low field strength, allowing the study of the Dirac vacuum.
\end{abstract}


\maketitle


\pagestyle{plain}



There has been a tremendous amount of efforts in the last few decades to increase laser intensities and it is now conceivable to reach laser intensities above $10^{23}$ W/cm$^{2}$ \cite{RevModPhys.78.309}, such that the corresponding field strength 
is comparable to Coulomb fields in matter. For slightly lower intensities, this has led to the development of nonperturbative models of ionization such as tunnelling ionization in atomic physics \cite{RevModPhys.72.545} and Charge Resonance Enhanced Ionization (CREI) in molecular physics \cite{PhysRevA.52.R2511}. The new regime of laser intensities now available allows the study of new physical effects where relativistic and Quantum Electrodynamics (QED) corrections start to be important \cite{Salamin200641}. Among the QED effects, one of the most important phenomenon is the long-sought Schwinger's mechanism \cite{PhysRev.82.664}, which consists of the decay of the vacuum of a static electric field into electron-positron pairs. In the Dirac interpretation, this can be seen as a tunneling of electrons from the negative energy sea to the positive continuum. This has never been observed experimentally because it requires field strength on the order of $E_{S} \sim 10^{18}$ V/m, which are not available experimentally (it corresponds to a laser intensity of $10^{29}$ W/cm$^2$). However, given the new experimental advances and the novel laser technologies, there has been a renewed interest in this process and new ideas have emerged which could allow to probe the QED vacuum. Therefore, many variations of Schwinger's original idea using different field configurations were proposed \cite{PhysRevD.2.1191,Nikishov1973,PhysRevE.66.016502,PhysRevLett.102.080402,PhysRevA.67.063407,PhysRevLett.100.010403,PhysRevLett.91.223601,PhysRevA.85.033408,PhysRevD.80.111301,PhysRevLett.104.250402}. A semi-classical theory of relativistic tunneling ionization has also been considered \cite{PhysRevLett.89.193001}. 

\begin{figure}
\includegraphics[width=0.8\textwidth]{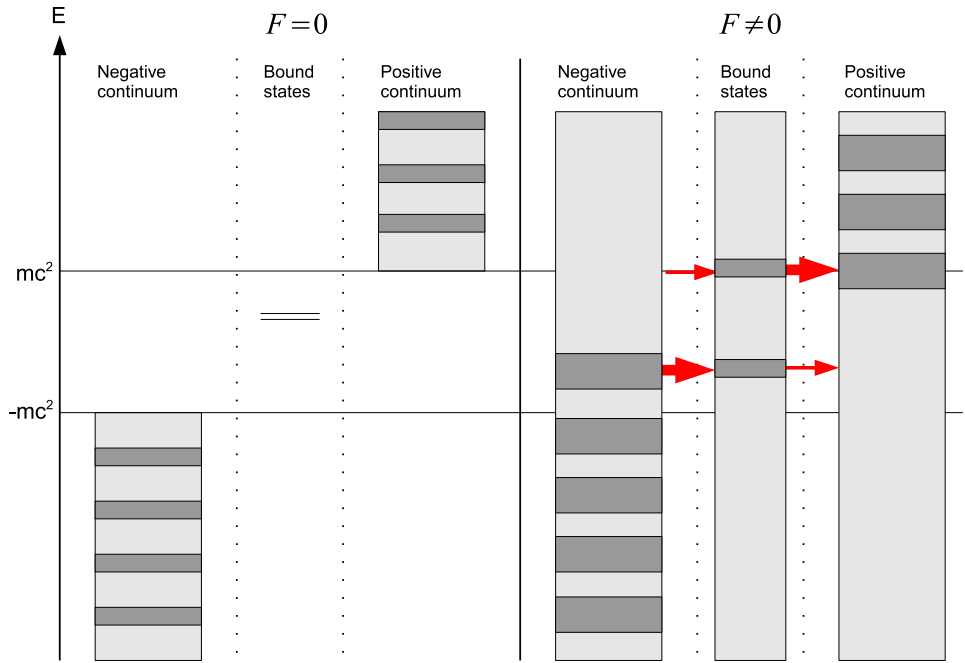}
\caption{Description of the REPP mechanism. The dark grey regions are the position of resonances (high density of states) while the light grey regions correspond to accessible energies which have a lower density of state.}
\label{fig:REPP}
\end{figure}

In this letter, a new mechanism is proposed to enhance pair production from a laser field interacting with a fully stripped diatomic molecule, which we denote the Resonantly Enhanced Pair Production (REPP) process. This mechanism is analogous to the CREI process, which is well-known in molecular physics \cite{PhysRevA.52.R2511}. It proceeds in the following way and is depicted in Fig. \ref{fig:REPP}. First, let us consider the case when the electric field $F$ is zero and look at the spectrum of the Dirac operator for a system having two nuclei. Also, for simplicity, we consider only two bound states: the ground state and the first excited state. If the nuclear charge is not too large ($Z<137$), the bound states are in the mass gap, in the energy range $[-mc^{2},mc^{2}]$. On the other hand, the negative and positive energy continua incorporate the so-called Ramsauer-Townsend resonances (RTR) (these are shown in dark on Fig. \ref{fig:REPP}), related to the backscattering of waves on the potential wells. This results in a peak-valley structure in the spectral density function in both continua \cite{1751-8121-45-21-215304}. When the electric field is turned on and the interatomic distance $R$ is varied, the position of resonances in the complex energy plane changes: the RTR of the negative energy sea are pushed to higher energies while the RTR of the positive energy continuum are pulled towards lower energies \cite{1751-8121-45-21-215304}. The bound states become resonances (they gain an imaginary part and thus, becomes unstable states) and are Stark-shifted by $\Delta \sim \pm FR/2$. When a RTR of the negative energy sea crosses a resonance from the bound states or a RTR from the positive energy continuum, the transition from the negative to the positive energy continuum is enhanced, yielding a higher pair production rate (channel 1). The same mechanism occurs for the excited state: when it crosses a RTR from the positive energy sea, the ionization of the molecule is enhanced, facilitating also the transition from the negative energy sea to the resonance because it reduces the Pauli blocking (if the electron is ionized, the excited state resonance is ``free'' and can receive a new electron that tunneled from the negative energy state). This is the channel 2. Therefore, if the field or the interatomic distance is not too large, the pair production occurs by a two-steps process: electrons from the negative energy sea tunnel to one of the bound state resonances, which is then followed by an ionization process where the same electron tunnels to the positive energy continuum. This produces a flux of electron and positrons, propagating in opposite direction. There is another possibility which occurs when the RTR from the positive and negative continua cross each other (channel 3). Then, it is possible to have a direct transition from the negative to the positive energy continua, which also enhances pair production but which effect is usually more modest than other channels. These effects are important at large atomic distance. For small atomic distances, such as in those achieved in Heavy ion collisions, the wave functions of each nucleus overlap and another mechanism is responsible for pair production: the effective electric charge in this case approaches $2Z$ in which case the ground state has a lower energy level. In the electric field, it is then easier for an electron to tunnel from the negative energy sea and again, this can lead to enhanced pair production.   

To confirm these ideas, a calculation of the pair production rate is performed in a very simple 1-D model: the nuclei are modeled by Dirac potential wells while the laser field is considered in the adiabatic limit and is taken as a static electric field. In a previous 3-D model of the one electron H$_{2}^{+}$ molecule in superintense laser fields, it was shown how to adapt Volkov solutions of the Dirac equation to the two centre problem \cite{0953-4075-42-17-171003}. In the present work, we use a simpler model which allows us to understand the mechanisms and the physics behind the pair production from the interaction of lasers with fully stripped diatomic molecules. We show that REPP can be utilized to enhance pair production at low field strength and large internuclear distance. This conclusion is reached by conducting a comparative study with the cases where there is no nuclei (Schwinger's process) and where there is only one nucleus.



\begin{figure}
\includegraphics[width=0.8\textwidth]{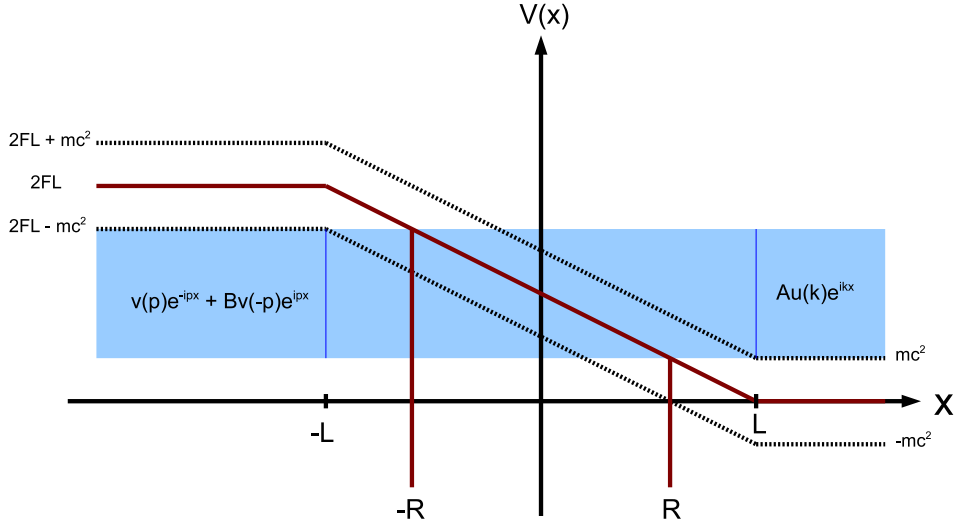}
\caption{Simple model to study REPP. The electric field has a finite extent in space. In blue is the Klein region where it is possible to have a transition from a negative to positive energy states. }
\label{fig:pair_prod_calc}
\end{figure}

To calculate the average rate of pair produced, we follow the discussion presented in \cite{1977mgm..conf..459D,PhysRevD.38.348,1402-4896-23-6-002} and assume that the electric field vanishes at $x=\pm \infty$ and thus, that it has a finite extent in space. In this case, it is possible to define the ``asymptotic states'' at $x=\mp \infty$: in these regions, the particles are free and there is a natural separation between the negative and positive energy solutions. This allows us to evaluate the number of pairs produced from a solution of the Dirac equation. It should be noted here that boundary conditions on the wave function at $x=\mp \infty$ are obtained from the time-dependent case where the Lehmann-Symanzik-Zimmermann (LSZ) asymptotic conditions, which implies the vanishing of the field at $t=\mp \infty$, can be used \cite{Baltz2001395}. In the time-independent case however, the latter cannot be fulfilled directly: we have to consider localized wave packets which are effectively in the free region when $t=\mp \infty$. From these considerations, it is possible to evaluate the average number of pairs $\langle n \rangle$ and it has been argued in \cite{1977mgm..conf..459D,PhysRevD.38.348,1402-4896-23-6-002} that this observable is given by
\begin{eqnarray}
\frac{d \langle n \rangle}{dtdE} = \frac{1}{2\pi} |A(E)|^{2},
\end{eqnarray}
where $A$ is the coefficient of the positive energy solution propagating towards $x=+\infty$, at the right of the potential (see Fig. \ref{fig:pair_prod_calc}). This formula is valid for a time-independent external field where solutions are labeled by energy. Similar formula have been derived in \cite{PhysRevD.82.025015,PhysRevD.80.065010}.

Thus, the calculation of pair production reduces to a transmission-reflection problem where the incident, reflected and transmitted waves are given respectively by:
\begin{eqnarray}
\psi_{\rm inc.}(x) &=& v(p)e^{ip(E)x} ,\\
\psi_{\rm ref.}(x) &=& Bv(-p)e^{-ip(E)x} ,\\
\psi_{\rm trans.}(x) &=& Au(k)e^{ik(E)x} .
\end{eqnarray}
Here, $u,v$ are the positive and negative energy free spinors: their explicit expression will be given below.

We are considering the 1-D Dirac equation, which is given by
\begin{eqnarray}
\label{eq:dirac_eq_ti}
E \psi(x) = \left[ -ic\sigma_{z} \partial_{z} + \sigma_{x} mc^{2} + A_{0}(x) + V(x)  \right] \psi(x),
\end{eqnarray}
where $E$ is the eigenenergy, $c$ is the light velocity, $m$ is the electron mass and $\psi(x)$ is the two-components spinor wave function. The electric potential is divided in three spatial regions as (see Fig. \ref{fig:pair_prod_calc})
\begin{eqnarray}
A_{0}(x) = 
\begin{cases}
2FL & \mbox{for} \; x \in (-\infty,-L] \\
-F(x-L) & \mbox{for} \; x \in (-L,L) \\
0 & \mbox{for} \; x \in [L,\infty) 
\end{cases}
\label{eq:elec_pot}
\end{eqnarray}
where $F$ is the field strength (we are working in a gauge where the vector potential is $A_{x}=0$ and the electric field is related to the potential as $E_{x}=-\partial_{x}A_{0}(x)$) and $2L$ is the length over which the electric field is constant. Outside the interval $[-L,L]$, the electric field vanishes. 

For the other scalar potential, three cases are considered:
\begin{enumerate}
	\item No nucleus: $V(x)=0$
	\item Single nucleus: $V(x)=-g\delta(x)$
	\item Two nuclei: $V(x) = -g \delta(x-R) -g \delta (x+R)$
\end{enumerate}
where $g$ is the potential strength (physically, it is related to the charge of the nucleus). It was shown in \cite{1751-8121-45-21-215304} that the Dirac delta potential wells can be characterized by the following boundary conditions: $\lim_{\epsilon \rightarrow 0}\psi(\zeta + \epsilon) = \lim_{\epsilon \rightarrow 0}G\psi(\zeta-\epsilon)$ ($\zeta$ is the potential well position), which relates the wave function on the right and the left of the potential well. The matrix components are given by $G_{12} = G_{21}=0$ and
\begin{eqnarray}
\label{eq:bound_cond1}
G_{11} &=& \frac{1}{1+\frac{g^{2}}{4c^{2}}} \left[ 1-\frac{g^{2}}{4c^{2}} + i \frac{g}{c} \right]=G_{22}^{*}  .
\end{eqnarray}
%
%
%
%
Thus, the wave function between the potential wells is a solution of Eq. \eqref{eq:dirac_eq_ti} without $V(x)$. 
%
%
Then, Eqs. \eqref{eq:bound_cond1} is used to match the wave function at $x=0,\pm R$, for case 2 and 3 respectively.

The Dirac equation can be solved analytically by decoupling the two spinor components and by letting $ y(x) = e^{-i\frac{\pi}{4}}\sqrt{\frac{2c}{F}} \left( \frac{E-F(x-L)}{c} \right)$. 
%
%
Then, the Dirac equation becomes a system of equations with well-known solutions in terms of parabolic cylinder functions $U(\gamma,z)$ \cite{DLMF}:
\begin{eqnarray}
 \psi(x) &=& c_{1} U_{a}(x) + c_{2} U_{b}(x),
\end{eqnarray}
%
where $c_{1,2}$ are integration constants and where we defined
\begin{eqnarray}
U_{a,1}(x)&\equiv & U(\gamma,y(x)) ,\\ 
U_{b,1}(x) &\equiv &  U(-\gamma,-iy(x)), \\
U_{a,2}(x)&\equiv &mc \sqrt{\frac{c}{2F}} e^{i\frac{3\pi}{4}} U(\gamma+1,y(x)) ,\\
U_{b,2}(x) &\equiv & \frac{1}{mc} \sqrt{\frac{2F}{c}} e^{-i\frac{\pi}{4}} U(-\gamma-1,y(x)) .
\end{eqnarray}
Here, we have $\gamma = i\frac{m^2 c^3}{2F} - \frac{1}{2}$.
The last missing ingredient to compute pair production are the negative and positive energy free spinors. There is a well-known procedure to compute these quantities where one seek plane wave solutions. This yields
\begin{eqnarray}
 u(k) &=& \frac{1}{\sqrt{2E}}
\begin{bmatrix}
 \sqrt{E+ck(E)} \\
 \sqrt{E-ck(E)}
\end{bmatrix}, \\
%
%
v(p) &=& \frac{1}{\sqrt{2(E-2FL)}}
\begin{bmatrix}
 \sqrt{(E-2FL)+cp(E)} \\
- \sqrt{(E-2FL)-cp(E)}
\end{bmatrix}
\end{eqnarray}
where $k(E)=\frac{1}{c}\sqrt{E^{2}-m^{2}c^{4}}$ and $p(E)=\frac{1}{c}\sqrt{(E-2FL)^{2}-m^{2}c^{4}}$. 


Now, imposing the continuity of the wave function at the region boundaries and using Eq. \eqref{eq:bound_cond1} to include the potential wells, we obtain the following conditions for the three cases:
\begin{enumerate}
	\item No nucleus:
\begin{eqnarray*}
v(p)e^{ipL} + B v(-p)e^{-ipL} =  a_{1} U_{a}(-L) + a_{2} U_{b}(-L), \\
a_{1} U_{a}(L) + a_{2} U_{b}(L) = Au(k)e^{ipL} .
\end{eqnarray*}
\item Single nucleus:
\begin{eqnarray*}
v(p)e^{ipL} + B v(-p)e^{-ipL} =  a_{1} U_{a}(-L) + a_{2} U_{b}(-L), \\
a_{1} U_{a}(0) + a_{2} U_{b}(0) = G^{-1}\left[b_{1} U_{a}(0) + b_{2} U_{b}(0) \right], \\
b_{1} U_{a}(L) + b_{2} U_{b}(L) = Au(k)e^{ipL} .
\end{eqnarray*}
\item Two nuclei:
\begin{eqnarray*}
v(p)e^{ipL} + B v(-p)e^{-ipL} =  a_{1} U_{a}(-L) + a_{2} U_{b}(-L), \\
a_{1} U_{a}(-R) + a_{2} U_{b}(-R) = G^{-1}\left[b_{1} U_{a}(-R) + b_{2} U_{b}(-R) \right] ,\\
b_{1} U_{a}(R) + b_{2} U_{b}(R) = G^{-1}\left[c_{1} U_{a}(R) + c_{2} U_{b}(R) \right], \\
c_{1} U_{a}(L) + c_{2} U_{b}(L) = Au(k)e^{ipL} .
\end{eqnarray*}
\end{enumerate}
These are systems of equations which can be used to solve for the integration constants $A,B,a_{1,2},b_{1,2},c_{1,2}$. The numerical results are presented in the following.

\begin{figure}
\begin{flushleft}
\includegraphics[width=0.88\textwidth]{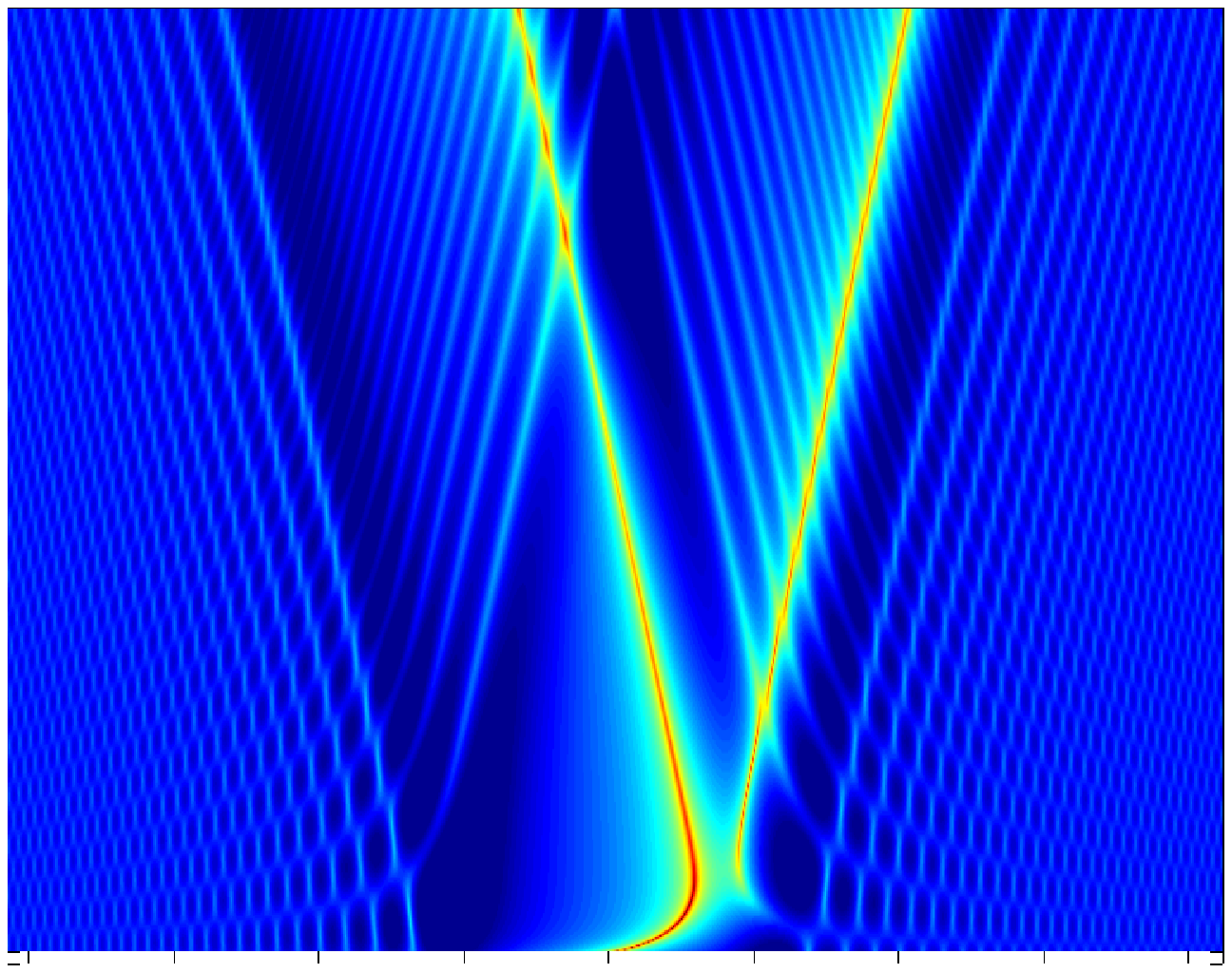}
\end{flushleft}
\vspace{-4.67in}
\begin{flushleft}
\includegraphics[width=0.88\textwidth]{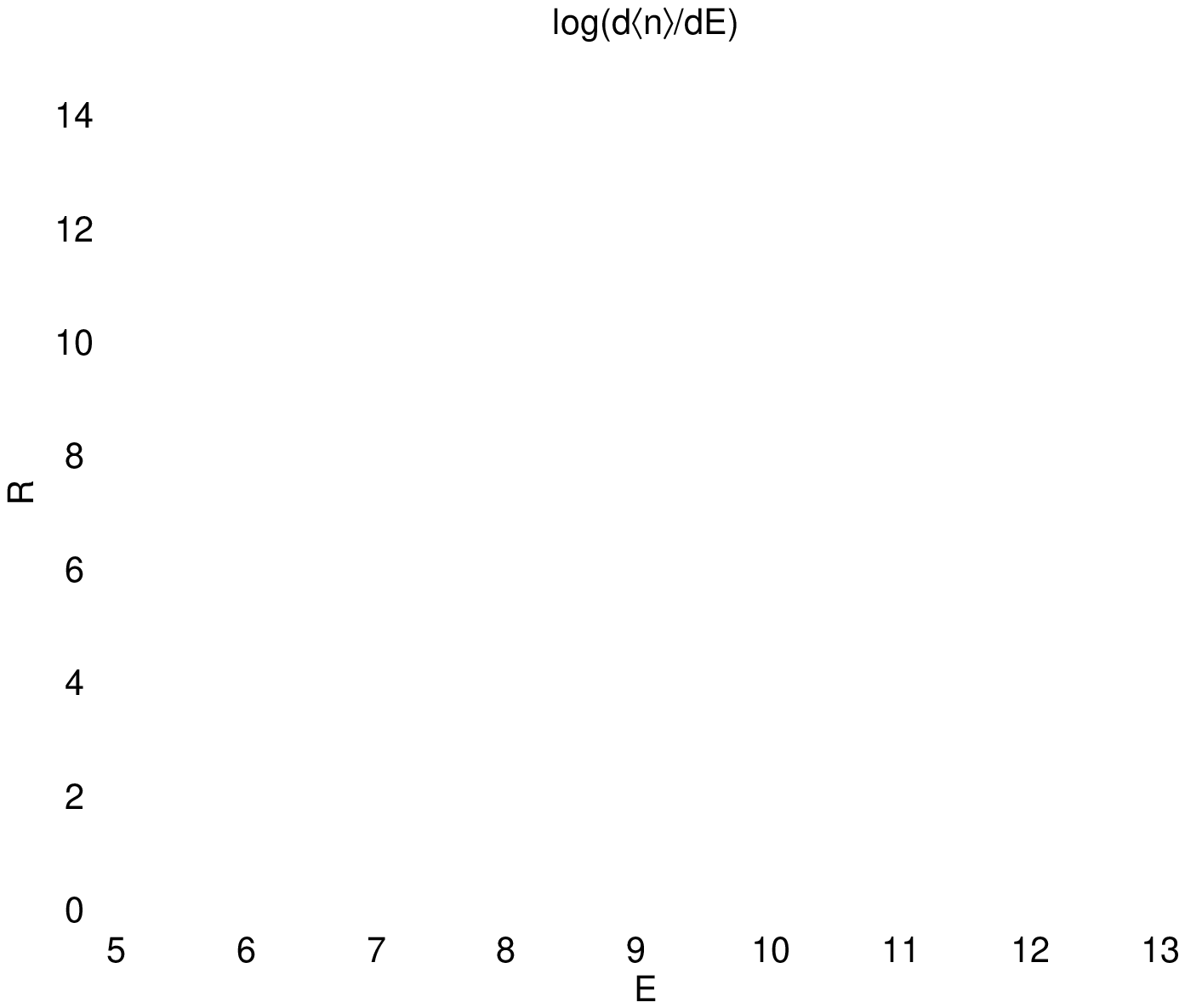}                                                  
\end{flushleft}
\vspace{-4.63in}
\begin{flushright}
\includegraphics[width=0.88\textwidth]{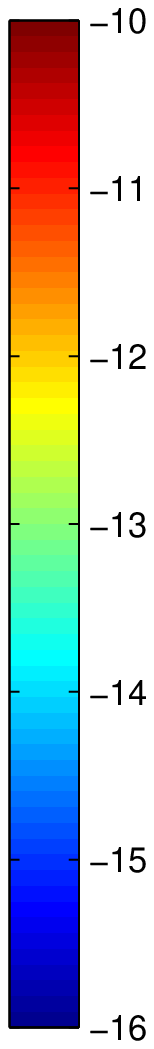}      
\end{flushright}
\caption{Numerical results for $d\langle n \rangle/dEdt$, for $g=0.8$, $F=0.09$ and $L=100.$. }
\label{fig:dnde_g1}
\end{figure}

The particle spectrum $d\langle n \rangle/dEdt$ as a function of interatomic distance is plotted in Fig. \ref{fig:dnde_g1} for the two nuclei case and for $g=0.8$ (the ground state energy of this potential well corresponds to the energy of the 1s orbital of the U$^{91+}$ ion). The other parameters are chosen as $L=100.$, $F=0.09$, in units where $\hbar=c=m=1$ and $e=\sqrt{\alpha}$. In this figure, there is a clear enhancement of pair production at the position of resonances. Moreover, as $R$ is varied, it is possible to see how the resonances are moving in the energy plane: some of the RTR in the negative energy sea are moving to higher energies while the ground state and the first excited state are Stark shifted. More interesting is the fact that when the RER crosses with the bound states, there is an enhancement of pair production (channel 1 or 2). 

\begin{figure}
\includegraphics[width=0.8\textwidth]{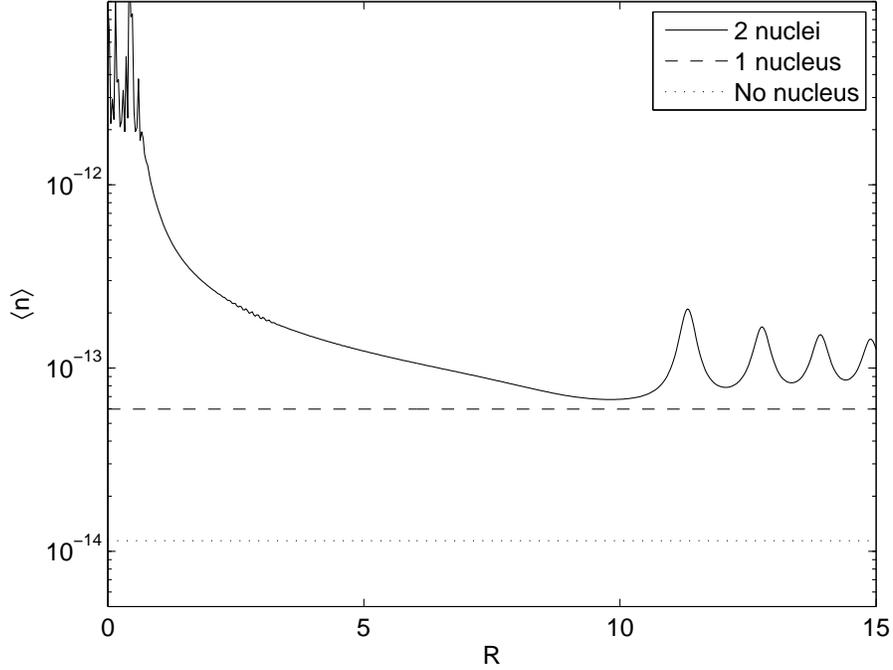}
\caption{Numerical results for $d\langle n \rangle/dt$. The oscillations at small $R$ are numerical artifacts.}
\label{fig:n_g1}
\end{figure}

In Fig. \ref{fig:n_g1}, the total rate $d\langle n \rangle/dt$ as a function of the interatomic distance is presented and compared to the other cases. This figures show clearly that the number of pairs is enhanced by the REPP at larger $R$: the position of peak in the rate corresponds to the interatomic distance where the ground state resonance crosses with RTR (channel 1). Thus, the number of pairs is approximately an order of magnitude higher than other cases and therefore, REPP is an important process for pair production in laser-matter interaction. The largest enhancement however occurs at small $R$ where it reaches two order of magnitude above the single nucleus case. This suggests that an experiment using HIC in a laser field could also be used to probe the Dirac vacuum.

\bibliographystyle{apsrev}

\bibliography{bibliography}

\end{document}